\documentclass[man,donotrepeattitle,12pt,floatsintext]{apa7}

\usepackage[american]{babel}

\usepackage{hyperref}
\usepackage{amsmath}
\usepackage{cleveref}
\usepackage{tabularx}
\usepackage[T1]{fontenc} 
\usepackage{helvet}      
\usepackage{lineno}

\usepackage{listings}
\usepackage{csquotes} 
\usepackage[style=apa,sortcites=true,sorting=nyt,backend=biber]{biblatex}
\DeclareLanguageMapping{american}{american-apa} 
\title{Relationships Between Trust, Compliance, and Performance for Novice Programmers Using AI Code Generation}
\shorttitle{NOVICE TRUST IN AI CODE GENERATION} 
\authorsnames[1,1,1]{Nicholas Gardella,Matthew L. Bolton,Sara L. Riggs}
\authorsaffiliations{{Department of Systems and Information Engineering, University of Virginia}}

\note{\subsection*{Précis}
A study of 27 novice programmers using the GitHub Copilot AI-driven Development Environment explored correlations between pre-interaction trust, AI compliance, programming performance, and post-interaction trust. The results could not show a relationship from trust to compliance but did show positive relationships from compliance to performance and performance back to trust.
}

\authornote{
    This work was supported by the University of Virginia Distinguished Fellowship, 
    NSF National Research Traineeship for Cyber-Physical Systems (1829004), 
    Graduate Research Fellowship Program for Computer and Information Science 
    and Engineering (1842490), and the Virginia Commonwealth Cyber Initiative 
    Central Virginia Node (GR102900). We have no conflicts of interest to disclose.
    Correspondence concerning this article should be addressed to Nicholas Gardella, 
    Olsson Hall, 151 Engineer's Way, Charlottesville, VA 22904, United States. 
    Email: njg4ne@virginia.edu
}

\abstract{
\subsection*{Objective}\label{sec:trust-abstract-objective}

To explore how novice programmers' trust in Artificial Intelligence-driven Development Environments (AIDEs) relates to their coding performance and AI compliance while programming under time pressure.

\subsection*{Background}\label{sec:trust-abstract-background}

Computer programming has undergone rapid upheaval due to state-of-the-art AIDEs, which provide clever automation for many aspects of software development. A longstanding interest of researchers of automation more generally has been the attitude of trust. Decades of research seek to explain how influencing trust can help to achieve desirable outcomes in different domains, but very limited work has provided similar focus on trust in AIDEs.

\subsection*{Method}\label{sec:trust-abstract-method}

We collected subjective measures of trust along with objective measures of performance and AIDE compliance from a diverse group of 27 novice programmers between two study locations.

\subsection*{Results}\label{sec:trust-abstract-results}

Our results corroborated traditional understandings of how trust changes through experiences. However, we did not find a relationship between trust and subsequent compliance during programming tasks. Greater compliance was associated with strong performance, and strong performance led to greater subsequent trust.

\subsection*{Conclusion}\label{sec:trust-abstract-conclusion}

Our findings raise new questions about the utility of trust in the context of interacting with AIDEs and generative AI. We call for further research into the effect of trust on compliance to recommendations from imperfect AI.

\subsection*{Application}\label{sec:trust-abstract-application}

This work can inform the design of training and educational content for generative AI use within and beyond software development. Instructional designers should consider risks of AI misuse and disuse and focus on promoting desirable interaction outcomes, regardless of trust's connection to them.
}


\begin{document}
\maketitle 

\iftrue
\section{Introduction}\label{sec:trust-introduction}

Artificial Intelligence (AI) technology can be very effective within software development, a field which drives technological progress in a myriad of application domains. AI-driven Development Environments (AIDEs) can help to automate aspects of coding such as typing, translating, problem-solving, testing, and debugging. AIDEs typically offer chat messaging, interactive text editing, or both. Programmers can use AIDEs as addons to their existing Integrated Development Environments to generate new text, edit existing text, learn more about a topic, or take an action (e.g., run a terminal command). AIDEs can produce outputs that are fast, highly effective, and well-liked, but can also produce incorrect or unsafe code \cite{gardella_performance_2024,pearce_asleep_2022,vaithilingam_expectation_2022,weisz_perfection_2021}. As such, it is important to study how programmers can use AIDEs to boost their productivity and well-being without compromising the safety of anyone their software will impact \cite{hoff_trust_2015,lee_trust_2004,lindsey_biology_2025,sadowski_rethinking_2019}. Unfortunately, it is not yet possible to ensure that AIDEs will effectively balance these goals \cite{lindsey_biology_2025}. Instead, human programmers must decide whether and to what extent they should rely on AIDEs. According to the human-automation interaction literature, this decision is largely based on a user's \emph{trust} in an automated system \cite{hoff_trust_2015,lee_trust_2004,passi_overreliance_2022}. With this connection, ensuring trust aligns closely with reliability should lead to ideal reliance.

This work focuses on relationships between trust, compliance, and performance in the context of novice programmers using AIDEs. Novices, as compared to intermediate and advanced programmers, have limited programming expertise to bias their attitudes, intentions, and behaviors. They are ideal candidates for this work because their higher openness increases the likelihood of detecting any true relationships that might exist between trust, compliance, and performance. These three constructs have been central to several theoretical models of human-automation interaction \cite{hoff_trust_2015,lee_trust_2004,mayer_integrative_1995}, but have not been studied extensively in programmer-AI interactions. These constructs are important because they are measurable, interconnected, and either intrinsically desirable or responsive to intervention. Lee and See \cite{lee_trust_2004} can be credited for popularizing the notion that trust, compliance, and performance interact cyclically, informing many later models \cite{rodriguez_rodriguez_review_2023}. Performance is desirable because it is the purpose of the automation to begin with. Compliance and reliance can help optimize the automation's effect on performance. Finally, interventions can target trust as an avenue for optimizing compliance and thus performance. However, the utility of trust is limited if these connections break down. Therefore, the goal of this work is to evaluate the relationships between novice programmers' (a) trust in an AI code generator (b) compliance with the AI's suggestions, and (c) programming performance.

\subsection{Trust, Compliance, and Performance}\label{sec:trust-trust-compliance-and-performance}

Given the nascence of AIDEs, there is limited existing work about how trust, compliance, and performance are related for programmers using AI. However, the broader knowledge base around human-automation interaction supports a cyclic progression between these three constructs. Trust is thought to inform compliance, which, based on reliability, impacts performance, and trust is then adjusted based on performance \cite{hoff_trust_2015,lee_trust_2004,mayer_integrative_1995}. \Cref{fig:trust-cycle} captures this expected cycle, which was crystallized by \cite{lee_trust_2004}.

\begin{figure}[htbp]
\caption{Expected cycle of trust, compliance, and performance from automation trust literature.}
\label{fig:trust-cycle}
{
    \setkeys{Gin}{alt={A diagram shows a start at Trust (a) with a dashed arrow to Compliance (b), which has a dashed arrow to Performance (c), which has a dashed arrow back to Trust (a).}}
    \includegraphics[width=.75\linewidth]{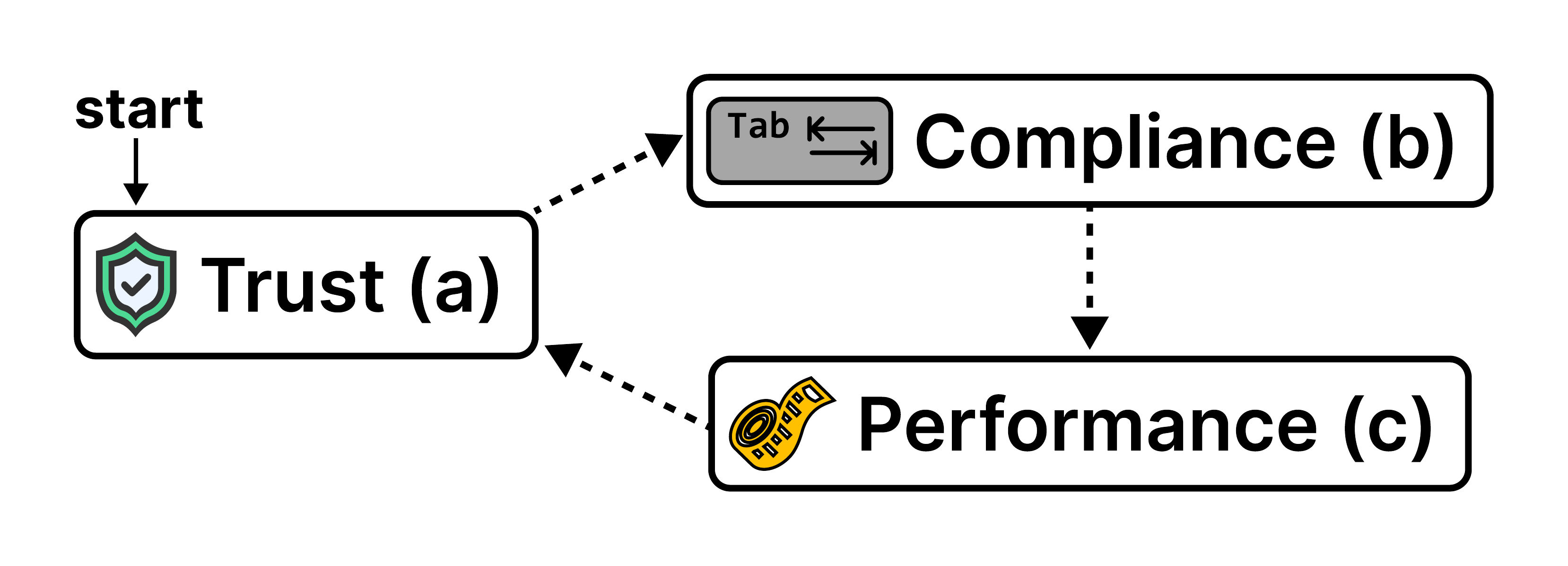}
}

\noindent
\textit{Note.} Users start with an initial level of (a) trust in the AIDE. Next, trust informs (b) how much users comply with AIDE suggestions. When the AIDE is very reliable for the task at hand, this compliance impacts (c) performance. Depending on how the team performs, the user can reevaluate their (a) trust.
\end{figure}

\emph{Note}. Users start with an initial level of (a) trust in the AIDE. Next, trust informs (b) how much users comply with AIDE suggestions. When the AIDE is very reliable for the task at hand, this compliance impacts (c) performance. Depending on how the team performs, the user can reevaluate their (a) trust.

In addition to theoretical evidence, empirical studies support that trust can predict compliance and be predicted by performance. Quantitative evidence for trust predicting compliance comes from the automation trust literature \cite{daly_task_2002,merritt_not_2008,patton_relationship_2024,wetzel_driver_2005}, but many qualitative studies of AIDEs have suggested this connection as well \cite{jayagopal_exploring_2022,perry_users_2023,ross_programmers_2023,vaithilingam_expectation_2022}. As for performance predicting trust, quantitative evidence is also from the automation literature, but AIDE-related studies have supported this connection, too \cite{brown_identifying_2024,noauthor_investigating_nodate}. The following subsections further detail the overlap between empirical work and the theoretical underpinnings of a trust-compliance-performance cycle.

\subsubsection{Trust}\label{sec:trust-trust}

Human \emph{trust} in automation (\Cref{fig:trust-cycle}a) is generally considered an attitude one actor has toward relying on another whose reliability is not fully knowable \cite{ajzen_understanding_1980,lee_trust_2004,mayer_integrative_1995}. Trust varies in its generality---i.e. in whom or under what circumstances does one trust \cite{hoff_trust_2015,kramer_trust_1999,marsh_role_2003,mcknight_what_2000,merritt_not_2008}. Here we focus on one's specific trust in an AIDE at several points in time. After the trustor becomes familiar with the trustee, but before any substantial interactions, they can provide a baseline level of specific trust \cite{biros_effect_2003,hoff_trust_2015,merritt_not_2008}. Thereafter, specific trust is "history-based" \cite{kramer_trust_1999,lee_trust_1992,merritt_not_2008} because accumulated interactions can be taken into consideration.

Prior work has measured trust in AIDEs, but only once per participant. Amoozadeh et al. \cite{amoozadeh_trust_2024} found that students in an earlier CS course trusted AIDEs more than those in a later course at the same school. Similarly, Perry et al. found a trend for participants who wrote more secure code to trust an AIDE less than those who wrote less secure code \cite{perry_users_2023}. These trends may be attributable to automation bias, which causes users to over-value salient automated cues, even when more reliable information is available \cite{mosier_automation_1999}. Novices, especially, may over-trust AIDEs by naively comparing their suggestions to more reliable feedback mechanisms like compiler messages \cite{prather_its_2024}. Novices have also been found to trust AIDEs more for code comprehension than for code completion, perhaps because of better trust affordances \cite{shah_students_2025} or because of less perceived personal risk \cite{mosier_automation_1999}. We are not aware of any studies taking repeated measures of trust in AIDEs, nor testing its effect on compliance.

\subsubsection{Compliance}\label{compliance}

\emph{Compliance} refers to acting on the advice of automation \cite{dixon_independence_2007}. Specific trust is thought to influence humans' \emph{compliance} and \emph{reliance} on automation \cite{lee_trust_2004}, corresponding to the arrow between \Cref{fig:trust-cycle}a and \Cref{fig:trust-cycle}b; though, recent work questions the strength of this connection \cite{patton_relationship_2024,pharmer_transparent_2025}. Colloquially, terminology is used more loosely, but \emph{reliance} is strictly defined as the choice \emph{not to act} in the \emph{absence} of an automation alarm \cite{dixon_independence_2007,lee_trust_2004}. Therefore, the term \emph{compliance} (\Cref{fig:trust-cycle}b) better describes depending on an AIDE by accepting suggested code, edits, or actions \cite{dixon_independence_2007}. Empirical work supports a small positive relationship between pre-interaction trust in automation and compliance within the interaction \cite{patton_relationship_2024}. Studies have found this connection in simulations of military decision support \cite{daly_task_2002}, X-ray luggage screening tasks \cite{merritt_not_2008}, and multitask driving scenarios \cite{wetzel_driver_2005}. Though similar quantitative evidence is lacking in the human-AIDE interaction context, several qualitative studies have suggested that trust in an AIDE can influence compliance \cite{jayagopal_exploring_2022,perry_users_2023,ross_programmers_2023,vaithilingam_expectation_2022}.

On the other hand, factors such as high workload \cite{daly_task_2002,spain_effects_2009} and low operator self-confidence \cite{lee_trust_1992} can also lead to greater compliance and reliance. When a user is confident to rely on their own abilities, they have a lesser need to use automation \cite{dzindolet_role_2003}. In the previously mentioned study about AIDEs and code security, participants with more prior security experience tended to use the AI-generated code less \cite{perry_users_2023}. It is not clear whether they did so because of low trust in AI or simply high self-confidence.

The cause of AIDE compliance matters because numerous studies have claimed to show evidence of programmers over-relying on AIDEs \cite{kazemitabaar_how_2024,prather_its_2024,ross_programmers_2023,shah_students_2025,vaithilingam_expectation_2022}. Typically, over-reliance would be considered "misuse" \cite{parasuraman_humans_1997} of automation---a "poor calibration in which trust exceeds system capabilities" \cite{lee_trust_2004}. This aligns with studies observing participants accepting wrong or insecure code from an AIDE \cite{perry_users_2023,ross_programmers_2023}. In contrast, other observations have been labeled over-reliance because of the risk, not the presence, of a bad outcome. For example, researchers often consider accepting large amounts of AI generated code without engaging with it (i.e., carefully reading or editing it) to be over-reliance \cite{kazemitabaar_how_2024,shah_students_2025,vaithilingam_expectation_2022}. However, if the code perfectly satisfies its intended need, the trust literature might consider such behavior to be appropriate reliance. One way to measure a bad outcome objectively is programming performance.

\subsubsection{Performance}\label{sec:trust-performance}

\emph{Performance} is how well the human-automation team collectively performs the task at hand. In software development, it can be measured in many ways (e.g. readability, time complexity, or reliability), but for novices, the most relevant measures assess basic software functionality under time constraints \cite{mccracken_multi-national_2001}. For example, novice programming performance can be quantified as the duration required to produce a program that passes an objective set of tests, or the number of tests passed within a specified period.

In general, automation compliance improves performance when the automation is reliable for the task or degrades it otherwise \cite{hoff_trust_2015,lee_trust_2004}. AIDEs are very reliable for the types of simple programming tasks in which novices engage \cite{chen_evaluating_2021,finnie-ansley_my_2023}. Novices also perform better with AIDE assistance than alone, so compliance should be positively related to performance, at least to some limit \cite{fan_impact_2025,gardella_performance_2024,kazemitabaar_studying_2023}. This corresponds to the arrow between \Cref{fig:trust-cycle}b and \Cref{fig:trust-cycle}c.

Performance can be used to adjust subsequent specific trust \cite{lee_trust_2004}. The performance of the automation itself has been positively associated with trust in simulations of industrial automation \cite{lee_trust_1992}, military decision support \cite{dzindolet_role_2003}, and multitask driving scenarios \cite{wetzel_driver_2005}. Extending these findings to the net performance of the human and the automation together, Merritt and Ilgen also found a direct relationship from performance to specific trust \cite{merritt_not_2008}. Unfortunately, no studies have validated this link for the AIDE context. However, at least two studies cite the perception of suggestion quality as an important factor for the development of trust in an AI code generator \cite{brown_identifying_2024,noauthor_investigating_nodate}. This suggests that performing well with an AIDE would increase a user's trust in it provided that they notice how well they are performing \cite{hoff_trust_2015}. This link corresponds to the arrow from \Cref{fig:trust-cycle}c back to \Cref{fig:trust-cycle}a, completing a cycle that can repeat over time.

\subsection{Research Questions \& Contributions}\label{sec:trust-research-questions-contributions}

While the evolving knowledge base on human-AIDE interaction acknowledges the relevance of trust, compliance, and performance, there is a major gap in its ability to describe their interrelationships empirically. Most quantitative evidence to date comes from studies with other types of automation. Our contribution is to empirically evaluate the cyclical trust-compliance-performance theory popularized by \cite{lee_trust_2004} in the context of human-AIDE interactions. To this end, we explore the following research questions in this study:

\begin{itemize}
\item
  \textbf{RQ1}: Does specific trust in an AIDE predict subsequent compliance (\Cref{fig:trust-cycle}, a to b)?
\item
  \textbf{RQ2}: Does compliance predict performance when coding with an AIDE (\Cref{fig:trust-cycle}, b to c)?
\item
  \textbf{RQ3}: Does programming performance predict subsequent specific trust in an AIDE (\Cref{fig:trust-cycle}, c to a)?
\end{itemize}

To answer these research questions, we planned to test three hypotheses using one regression coefficient test for each research question. First, for RQ1, we expected a positive relationship from specific trust (\Cref{fig:trust-cycle}a) to compliance (\Cref{fig:trust-cycle}b) \cite{daly_task_2002,merritt_not_2008,patton_relationship_2024,wetzel_driver_2005}. Due to the AIDE's high reliability on simple programming tasks like the ones we use \cite{chen_evaluating_2021,finnie-ansley_my_2023}, we also expected a positive relationship between compliance (\Cref{fig:trust-cycle}b) and performance (\Cref{fig:trust-cycle}c) for RQ2 \cite{hoff_trust_2015,klingbeil_trust_2024,lee_trust_2004,mcguirl_supporting_2006}. Finally, we expected a positive relationship from performance back to specific trust (\Cref{fig:trust-cycle}a, post-interaction) for RQ3 \cite{dzindolet_role_2003,lee_trust_1992,lee_trust_2004,merritt_not_2008,wetzel_driver_2005}.

\section{Method}\label{sec:trust-method}

Data was collected at University of Virginia (UVA) in Spring 2023 and Virginia State University (VSU) in Spring 2024. Performance data from UVA was previously reported for a separate analysis \cite{gardella_performance_2024}. This research complied with the American Psychological Association Code of Ethics and was approved by the Institutional Review Board at UVA for participants recruited there and at both universities for participants recruited at VSU. Informed consent was obtained from each participant.

\subsection{Participants}\label{sec:trust-participants}

In total, 27 participants were recruited by convenience as volunteers (16 female; 10 male; 1 unknown; 10 Black; 8 Asian; 3 Asian \& White; 4 White; 1 Latinx \& Black; 1 Latinx \& White). Only three participants had any prior experience with coding-specific AIDEs, but many (24) had tried general-purpose AI chatbots. Seventeen UVA students (11 female; 5 male; 1 unknown; 8 Asian; 3 Asian \& White; 4 White; 1 Latinx \& Black; 1 Latinx \& White) were recruited from a 1000-level Introduction to Programming (CS1) course. Ten VSU students (5 female; 5 male; 10 Black) were recruited from a 1000-level Introduction to Programming (CS2) course. By consulting syllabi and faculty at both universities, we determined that students in this second-semester intro course would have similar programming abilities to the participants from UVA. Nonetheless, we explicitly accounted for site-to-site variation throughout the analysis.

\subsection{Experimental Setup}\label{experimental-setup}

Participants at both sites programmed in the popular Visual Studio Code (VS Code) Integrated Development Environment on an Apple iMac with Python 3 and the GitHub Copilot IDE extension. At UVA, participants used GitHub Copilot circa 2023 (v1.79), while at VSU, participants used GitHub Copilot circa 2024 (v1.151) with GitHub Copilot Chat v0.11.1. This change kept VSU participants up to date with the newest supported versions available at the time and ensured that they did not encounter backwards compatibility issues with outdated remote services. Both versions provided inline suggestions and a multiple suggestions panel, but the newer version also had conversational options, context direction tools, and an interactive editor. Custom VS Code extensions provided task access, submission capabilities, and tracking of participants' interactions with Copilot. The interface showed a countdown timer, a PDF with task instructions, and a Python editor containing an unimplemented function. Participants could run and test their code using either the terminal or point-and-click options. They could advance to the next task in a trial only once their solution for the current task passed all the tests provided in the dataset---\href{https://github.com/openai/human-eval}{HumanEval} (https://github.com/openai/human-eval; \cite{chen_evaluating_2021}.

\subsection{Procedure}\label{sec:trust-procedure}

Participants at both sites were guided through the same procedure. They provided informed consent electronically and were given an overview of the study and an opportunity to ask questions. A series of training videos was presented to familiarize participants with the programming environment, the task submission tool, the functionality of GitHub Copilot, and the risks and benefits of using it. Because Copilot's functionality was expanded and the study software was updated slightly, participants at VSU were shown updated training videos on these topics to account for these updates. Participants at both sites viewed the same video about the risks and benefits of Copilot. While participants were shown different ways of using Copilot, they were not encouraged or discouraged from using any particular strategy. After each video, participants demonstrated their understanding either verbally or by completing the actions shown in the video.

As part of a larger study design, participants completed four Python programming blocks---two alone with standard IntelliSense autocomplete and two with GitHub Copilot. The original quantitative analysis of the UVA data provides additional details on the programming tasks participants worked on during these trials \cite{gardella_performance_2024}. In each block, participants saw up to eight tasks, where they could only view the next task after completing the current one. The ten participants at VSU were given identical tasks and AI treatments as ten of the participants at UVA. This study focuses only on the AI-assisted blocks and includes results using data from both sites. Trust data was not collected for solo trials as the IntelliSense autocomplete built into VS Code is deterministic and thus was not relevant to this study's research questions. As such, no comparison of trust in Copilot to a baseline of trust in IntelliSense was examined.

Before and after each AI-assisted block, participants completed a questionnaire containing a specific trust scale adapted from the work of Muir and Moray \cite{muir_trust_1996,muir_operators_1989}. For each trial, participants had 20 minutes to complete as many tasks as possible (up to eight). They were instructed to work for the best possible score, using the AI as much or little as needed. Bonus compensation was awarded to the top two scorers at each site to encourage adherence to this instruction. This was a key design choice for eliciting trust attitudes, as it created risk related to under-use or over-use of AI assistance during the time-pressured scenario \cite{mayer_integrative_1995,mosier_automation_1999}. \Cref{fig:trust-timeline} shows the timeline of specific trust measurements and programming tasks. Note that because participants also programmed alone as part of a larger counterbalanced design, specific trust measurements 1B and 2B were either taken in close succession (AI trials were back-to-back) or separated by one or two trials without AI. Counterbalancing means systematically varying the order of task sets used for each trial and the choice of which order to assign AI or solo conditions.

\begin{figure}[htbp]

\caption{Timeline of trust scale administration amidst AI-assisted Python programming.}
\label{fig:trust-timeline}
{
    \setkeys{Gin}{alt={A timeline shows training on Python and GitHub Copilot first, then Trust 1A, then 20 minutes of Coding with AI, then Trust 1B, then 0 to 2 of Solo Coding for 20 minutes, then the same series of trust, coding, and trust for trial 2.}}
    \includegraphics[width=\linewidth]{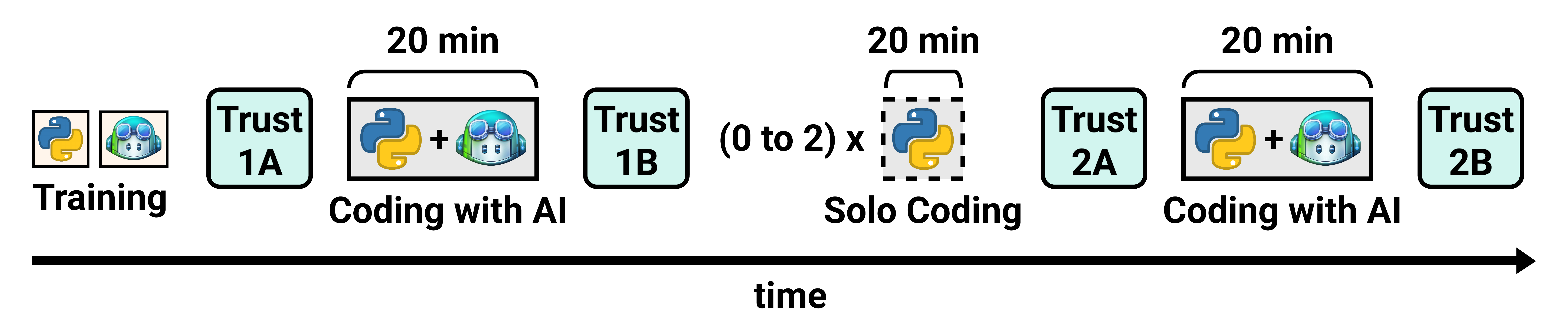}
}
\end{figure}

\subsection{Measures of Trust, Compliance, and Performance}\label{sec:trust-measures-of-trust-compliance-and-performance}

\subsubsection{Trust}\label{sec:trust-trust-1}

To measure trust in GitHub Copilot, we employed a specific trust scale adapted from the work of Muir \cite{muir_trust_1996,muir_operators_1989}. Muir and Moray concluded that perceptions of competence (see \Cref{eq:trust-muir-competence}), followed by predictability, dependability, and faith (see \Cref{eq:trust-muir-pred-dep-faith}) were key determinants of trust \cite{muir_trust_1996}. As shown in \Cref{table:trust-muir-scale}, we adapted both these four measurement items and the ground-truth overall trust item from \cite{muir_operators_1989} to our context. To avoid any ambiguity about the meaning of the word "trust" \cite{bolton_trust_2022}, this item was phrased as a perception of trustworthiness (reliability), which was clearly defined. The original scale was measured in millimeters on a paper instrument, so our digital version used a reasonable \cite{miller_magical_1956} zero to ten scale (0 = `\emph{Not At All'} or `\emph{Least Trustworthy'}, 10 = `\emph{Completely'} or `\emph{Most Trustworthy'}) with one point increments. \Cref{fig:trust-muir-sem-diagram} shows an SEM path diagram for the five observed trust items; the measurement model contains four items as indicators, and the factor they indicate ought to correlate with the fifth (summary) item. \Cref{eq:trust-sem} shows our application of \Cref{eq:trust-muir-competence,eq:trust-muir-pred-dep-faith} that handles the multicollinearity discussed in \cite{muir_trust_1996}.

    

\begin{gather}
    \text{(Overall Trust Item)} = 14.1 + 0.8 \cdot \text{Competence} + \epsilon \label{eq:trust-muir-competence} \\
    \textit{for residual } \epsilon \notag
\end{gather}

\begin{gather}
    \begin{split}
        \text{(Overall Trust Item)} = \beta_0 &+ \beta_1 \cdot \text{Predictability} + \beta_2 \cdot \text{Dependability} + \beta_3 \cdot \text{Faith} + \epsilon
    \end{split} \label{eq:trust-muir-pred-dep-faith} \\
    \textit{for coefficients } \beta \textit{ and residual } \epsilon \notag
\end{gather}

\begin{table}[htbp]
\caption{Subjective trust scale adapted from the Muir scale.}
\label{table:trust-muir-scale}
\begin{flushleft}
{
\renewcommand{\tabularxcolumn}[1]{m{#1}}
\begin{tabularx}{\textwidth}{l X}

\toprule
\textbf{Item} & \textbf{Wording} \\
\midrule

Competence & How confident are you in Copilot to help you solve problems with code? \\
Predictability & How confident are you for Copilot to behave predictably? \\
Dependability & How confident are you in Copilot to help you avoid errors and bugs in your code? \\
Faith & Assuming you have access to it, how confident are you in Copilot to help you with future programming for work or school? \\
Overall Trust & Trustworthiness is the extent to which GitHub Copilot is useful enough and effective enough to rely on to help someone program. How trustworthy do you think GitHub Copilot is? \\
\bottomrule
\end{tabularx}
}
\end{flushleft}
\noindent
\textit{Note.} The scale ranged from 0 to 10 with no decimals.
\end{table}

\begin{figure}[htbp]

\caption{SEM path diagram for the adapted Muir scale.}
\label{fig:trust-muir-sem-diagram}
{
    \setkeys{Gin}{alt={A path diagram starts at an overall trustworthiness observation and points with a regression label to a trust factor, which points with an indication label to the observations of competence, faith, dependability, and predictability.}}
    \includegraphics[width=.75\linewidth]{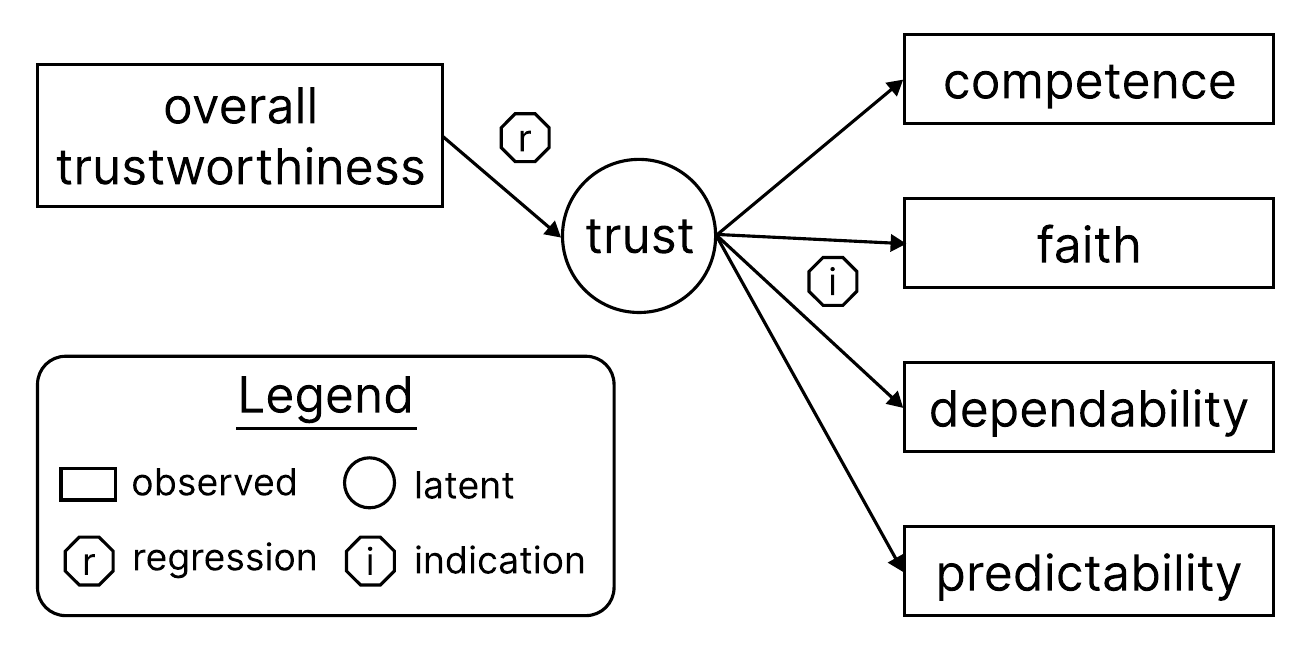}
}

\noindent
\textit{Note.} This Structural Equation Model (SEM) path diagram is composed of a four-indicator measurement model and an implied regression equation between an observed summary item (perceived trustworthiness) and the latent trust factor. We focused here on the measurement model only.
\end{figure}

\begin{gather}
    \text{(Trust Factor)} = \gamma_1 \cdot \text{(Overall Trust Item)} + \zeta \notag \\
    \begin{bmatrix} 
        \text{Competence} \\ \text{Predictability} \\ \text{Dependability} \\ \text{Faith} 
    \end{bmatrix} = 
    \begin{bmatrix} 
        \lambda_{\text{C}} \\ \lambda_{\text{P}} \\ \lambda_{\text{D}} \\ \lambda_{\text{F}} 
    \end{bmatrix} \cdot \text{(Trust Factor)} + 
    \begin{bmatrix} 
        \delta_{\text{C}} \\ \delta_{\text{P}} \\ \delta_{\text{D}} \\ \delta_{\text{F}} 
    \end{bmatrix} \label{eq:trust-sem} \\
    \textit{for coefficient } \gamma, \textit{ residual } \zeta, \textit{ loadings } \lambda, \textit{ and measurement errors } \delta \notag
\end{gather}


\subsubsection{Compliance}\label{compliance-1}

Compliance to GitHub Copilot's advice was operationalized by counting the total number of times that participants used built-in compliance features during each trial. This included accepting inline suggestions, shown as greyed-out text, or using a button in the multiple suggestions panel to accept one of its options. Notably, manual copy-and-pasting or typing of viewed suggestions was not counted. The newer version of Copilot in study two generated additional sources of compliances: (1) using buttons in the chat panel for copying or inserting provided code blocks and (2) accepting code edits from the inline chat widget. Once again, manual copy-pasting and typing were not counted. Only if the user clicked a built-in copy button was a copy tracked. It was important to count all sources of compliances from study two in case participants chose newer features in lieu of older ones.

\subsubsection{Performance}\label{sec:trust-performance-1}

Performance was measured using an ordinal score reflecting how many tasks a participant completed in 20 minutes (0-8 points). We were not sure about the measurement level of this data, as it is not possible to make tasks perfectly exchangeable in difficulty. Ultimately, we treated the score as a ratio-level count based on its nature and distribution. To ensure this assumption did not bias our conclusions, we also considered a binary transformation for all relevant regressions. That is, for the \nameref{sec:trust-conclusion-sensitivity-analysis}, score was recoded as either \emph{low} or \emph{high} against the median of all trials with AI across all 27 participants (2 completed tasks of a possible 8). A \emph{low} score was one at or below the median (0-2), while a \emph{high} score was one above the median (3-8).

\subsection{Data Analysis}\label{sec:trust-data-analysis}

All statistics were computed in R \cite{ihaka_r_1996}. Unless otherwise stated, we used the standard statistical significance threshold of $\alpha$ = .05.

\subsubsection{Scale Validation and Factor Analysis}\label{sec:trust-method-scale-validation-and-factor-analysis}

We assessed the internal consistency of the Muir scale using Cronbach's \cite{cronbach_coefficient_1951} $\alpha$. Next, we performed single-factor Confirmatory Factor Analysis (CFA) to assess unidimensionality. Goodness-of-fit for CFA models was assessed using an accepted spectrum of fit measures ($\chi^2$ $p$-value \textgreater{} .05, CFI \textgreater{} .95, SRMR \textless{} .08, RMSEA \textless{} .06) \cite{byrne_structural_2006,jackson_reporting_2009}. We compared two practically applicable CFA model structures using the Satorra-Bentler \cite{satorra_scaled_2001} scaled $\chi^2$ difference test. First, we loaded each item onto the factor equally to represent the common application of collapsing the item ratings from a trust scale with an arithmetic mean. Next, we allowed the item loadings to be estimated freely. We used the better model to compute trust factor scores for further analysis, and assessed test-retest reliability with a paired t-test and the Pearson correlation coefficient \cite{pearson_vii_1997}.

\subsubsection{Trust Score Computation}\label{sec:trust-trust-score-computation}

For computing trust scores, we applied the four-indicator CFA model using lavaan \cite{rosseel_lavaan_2012} with the \lstinline[language=R]{lavaan::lavPredict(type="ov")} function. This produced a score for each observed scale item response weighted by its contribution to the latent factor (determined by the CFA). We took row-wise means to produce latent factor scores distributed like the original data (not $\mu=0$, $\sigma^2=1$).

\subsubsection{Generalized Estimating Equations}\label{sec:trust-generalized-estimating-equations}

We explored the interrelationships of trust, compliance, and performance using Generalized Estimating Equations (GEE). GEE extends Generalized Linear Model (GLM) regression to relax the assumptions of observational independence and family-consistent residual dispersion by accounting for clustering during estimation \cite{liang_longitudinal_1986}. We also considered multi-level Bayesian regression, which aligns well to trust theory, but chose GEE as it is more robust and less sensitive to assumptions with as few clusters as our 27 \cite{mcneish_modeling_2016,pedroza_estimating_2017}. Compared to prior trust research that used pooled regression \cite{lee_trust_1992,merritt_affective_2011}, using GEE allowed us to explicitly account for dependence in our repeated-measures data.

Because we were not interested in using models for prediction, we were not concerned with the overall goodness-of-fit (i.e., R\textsuperscript{2} or Information Criterion values). Rather, we were interested in the significance of each relationship and interpreting the coefficients. Therefore, we used coefficient Wald tests to assess statistical significance according to regression coefficients and standard errors. To control Type I error, we used \lstinline[language=R]{geesmv::GEE.var.mbn}
\cite{wang_geesmv_2015} to compute Cluster Robust Standard Errors (CRSEs) via the Morel-Bokossa-Neerchal \cite{morel_small_2003} small-sample GEE correction \cite{mcneish_modeling_2014,mcneish_modeling_2016}. The coefficients themselves were estimated with \lstinline[language=R]{geepack::geeglm} \cite{hojsgaard_r_2006} with identical settings as geesmv. The working correlation structure was set to "exchangeable" with the two observations per cluster used for RQ1 and RQ2 but to autoregressive ("ar1" for geepack and "AR-M" for geesmv) with the four trust observations per cluster used for RQ3.

The theoretical model in \Cref{fig:trust-cycle} drove our choices for predictor and response measures, but we also included covariates for site and time. Trust scores were derived from the good-fit CFA models as described in Trust Score Computation and used as predictors for RQ1 and response measures for RQ3. Their symmetric distribution motivated a Gaussian family choice for RQ3. Compliances per trial were appropriately right skewed for a count process, motivating a Poisson family choice for RQ1. Performance score (at least ordinal; 0-8 points per trial) was right-skewed and treated as a Poisson process for RQ2. Given we used GEE and had more than seven discrete levels for both trust ratings and performance scores, relaxing assumptions about their measurement level was an ideal tradeoff over losing information and simplicity with an ordinal approach \cite{rhemtulla_when_2012}.

\subsubsection{Conclusion Sensitivity Analysis}\label{sec:trust-conclusion-sensitivity-analysis}

We considered the basic assumptions of GEE and the sensitivity of our conclusions (i.e., the direction and significance of our target predictors) to different reasonable assumptions. Though GEE is exempt from some regression assumptions, it is important to consider linearity, model specification, and influential observations. After choosing appropriate response distribution families, we confirmed general linearity by visual inspection. When specifying GLM formulas, we added covariates to account for time and research site. Next, we used \lstinline[language=R]{glmtoolbox::glmgee} \cite{vanegas_generalized_2023} and Cook's \cite{cook_detection_1977} distance (stats) to check our conclusions for sensitivity to possible influential observations. Given we omitted the "Overall Trust" summary item of the Muir scale (see \nameref{sec:trust-method-scale-validation-and-factor-analysis}) and used CFA-derived trust factor scores in a non-SEM analysis, we also checked for conclusion sensitivity to using a simple mean of all five items instead. Finally, we checked sensitivity to treating ordinal performance data as ratio-level (see \nameref{sec:trust-performance-1} and \nameref{sec:trust-generalized-estimating-equations}) using a binary transformation (i.e. at or below the median 0-2 points or above the median 3-8 points). This also called for a binomial family GEE when performance was the regression response. In each section of the \nameref{sec:trust-results} to follow, we comment on the sensitivity of our results with respect to these three considerations: influential observations, trust aggregation, and performance measurement level. Otherwise, the data met the assumptions of GEE.

\section{Results}\label{sec:trust-results}

\subsection{Scale Validation and Factor Analysis}\label{sec:trust-results-scale-validation-and-factor-analysis}

The Muir scale (see \Cref{table:trust-muir-scale}) produced Cronbach's $\alpha$ = .90 with the "ground truth" trustworthiness item included and $\alpha$ = .87 without it. This item was excluded from the measurement model on account of the covariance assumptions of CFA. The equal-weight four-indicator CFA fit poorly (Y-B $\chi^2$(6) = 56.96, \emph{p} \textless{} .001, CFI\textsubscript{R} \textgreater.730, RMSEA\textsubscript{R} = 0.305, SRMR = 0.381). A freely estimated model of the four indicators was significantly better (S-B $\chi^2$ Diff. = 57.351, \emph{p} \textless.001), and the overall fit was good (Y-B $\chi^2$(2) = 2.00, \emph{p} = .37, CFI\textsubscript{R} \textgreater.999, RMSEA\textsubscript{R} = 0.002, SRMR = 0.022). We used this model for computing trust scores at each timepoint. Participants had no new experiences with the AI between timepoints 1B and 2A (see \Cref{fig:trust-timeline}). Trust scores were test-retest reliable between these points according to the paired \emph{t}-test (t (26) = 0.520, \emph{p} = .61) and Pearson correlation (r (25) = .93, \emph{p} \textless{} .001).

\subsection{RQ1: Does specific trust in an AIDE predict subsequent compliance?}\label{sec:trust-rq1-does-specific-trust-in-an-aide-predict-subsequent-compliance-figure-1-a-to-b}

We first examined the relationship between the computed factor score for specific trust before a trial began (\Cref{fig:trust-timeline}, Trust 1A and 2A) and the number of times participants accepted text from Copilot using the built-in compliance features. With a Poisson family, pre-trial trust, along with covariates to control for trial number (1 v. 2) and research site (UVA v. VSU), was used as a predictor of compliance count. \Cref{table:trust-rq1-result} shows the results of the coefficient Wald tests based on the MBN CRSE correction. The exponentiation of the coefficients provides the multiplicative factor by which the models predict a single point of pre-trial specific trust (0 to 10) to scale the number of expected compliances during that trial. Pre-trial trust was not a statistically significant predictor of compliance, showing a slightly negative slope. The \nameref{sec:trust-conclusion-sensitivity-analysis} revealed no sensitivity to trust aggregation or influential observations. Of note, the trust coefficient's direction shifted from -1\% to +1\% when two influential clusters were excluded. There is no evidence that pre-trial trust was related to the number of times participants used the built-in compliance features. This finding calls into question the link in \Cref{fig:trust-cycle} between trust (a) and compliance (b). The estimated working correlation of \(\widehat{\text{$\alpha$}}\text{ = .83}\) supports the suspected observational dependence structure. The statistically significant covariates suggest higher compliance at UVA than at VSU and higher compliance on the second trial than on the first.

\begin{table}[htbp]
\caption{Exchangeable Poisson GEE estimates relating pre-trial trust to compliance count.}
\label{table:trust-rq1-result}
\begin{flushleft}
{
\begin{tabular}{l lllll}
\toprule
\multicolumn{1}{c}{\textbf{Variable}} & 
\multicolumn{1}{c}{\textbf{\textit{B}}} & 
\multicolumn{1}{c}{\textbf{\textit{e\textsuperscript{B}}}} & 
\multicolumn{1}{c}{\textbf{\textit{CRSE}}} & 
\multicolumn{2}{c}{\textbf{95\% CI}} \\

& & & & \multicolumn{1}{c}{\textit{LL}} & \multicolumn{1}{c}{\textit{UL}} \\ 
\midrule

Constant & 2.91$^{***}$ & $\hat{\mu}=\text{18}$ & 0.32 & 2.29 & 3.54 \\
Site = VSU & -0.87$^{***}$ & -58\% & 0.26 & -1.37 & -0.36 \\
Trial Number & 0.20$^{**}$ & +22\% & 0.07 & 0.06 & 0.33 \\
Pre-Trial Trust (A) & -0.01 & -1\% & 0.05 & -0.10 & 0.08 \\

\bottomrule
\end{tabular}
}
\end{flushleft}
\noindent
\emph{Note}. $k=27$ clusters (participants) of balanced size $n_i=2$ (trials per participant). We estimated the population-averaged relationship between the computed factor score for pre-trial trust and the number of compliances using Generalized Estimating Equations (GEE) with a Poisson link, exchangeable correlation structure, Morel-Bokossa-Neerchal (MBN) bias-corrected Cluster-Robust Standard Error (CRSE), and covariates for study site and time (i.e. trial number). The estimated working correlation was $\widehat{\text{$\alpha$}} = .83$.\\
$^{**}p < 0.01$; $^{***}p < 0.001$

\end{table}

\subsection{RQ2: Does compliance predict performance when coding with an AIDE?}\label{sec:trust-rq2-does-compliance-predict-performance-when-coding-with-an-aide-figure-1-b-to-c}

Pre-trial trust did not seem to relate to compliance, but we also wanted to see whether increased compliance led to any differences in performance. To test this, we again used Poisson GEE regression with an exchangeable correlation structure and the MBN correction. \Cref{table:trust-rq1-result} shows the results of the coefficient Wald tests. There was a small but statistically significant positive relationship between compliance and performance (\emph{p} \textless{} .01) corresponding to a roughly 2\% improvement for each additional compliance (\(\overline{\text{x}} \approx\) 19, SD \(\approx\) \(\widetilde{\text{x}} \approx\) 14). The significance of trial number suggests performance increased with additional experience regardless of the level of compliance. The \nameref{sec:trust-conclusion-sensitivity-analysis} revealed little sensitivity. Excluding two influential clusters increased the compliance coefficient from 2\% to 3\%, reducing uncertainty slightly. Dichotomizing score data caused the trial number covariate to lose statistical significance, a possible indicator of entanglement between compliance and experience. Compliance was still significant, increasing the odds of scoring above the median by a roughly 9:8 ratio per compliance event. \Cref{fig:trust-compliance-by-score} shows that participants were able to achieve high scores without correspondingly high levels of compliance, yet greater compliance clearly correlated with greater performance. The estimated working correlation of \(\widehat{\text{$\alpha$}}\text{ = .54}\) indicates some correlation between each participant's own scoring rates on the two trials.

\begin{table}[htbp]
\caption{Exchangeable Poisson GEE estimates relating compliance count to performance score.}
\label{table:trust-rq2-result}
\begin{flushleft}
{
\begin{tabular}{l lllll}
\toprule
\multicolumn{1}{c}{\textbf{Variable}} & 
\multicolumn{1}{c}{\textbf{\textit{B}}} & 
\multicolumn{1}{c}{\textbf{\textit{e\textsuperscript{B}}}} & 
\multicolumn{1}{c}{\textbf{\textit{CRSE}}} & 
\multicolumn{2}{c}{\textbf{95\% CI}} \\

& & & & \multicolumn{1}{c}{\textit{LL}} & \multicolumn{1}{c}{\textit{UL}} \\ 
\midrule

Constant & 0.18 & $\hat{\mu}=\text{1}$ & 0.29 & -0.39 & 0.74 \\
Site = VSU & 0.25 & +28\% & 0.30 & -0.35 & 0.84 \\
Trial Number & 0.34$^{**}$ & +41\% & 0.13 & 0.09 & 0.59 \\
Compliances & 0.02$^{**}$ & +2\% & 0.01 & 0.01 & 0.03 \\

\bottomrule
\end{tabular}
}
\end{flushleft}
\noindent
\emph{Note}. $k=27$ clusters (participants) of balanced size $n_i=2$ (trials per participant). We estimated the population-averaged relationship between the number of compliance events and the performance score using Generalized Estimating Equations (GEE) with a Poisson link, exchangeable correlation structure, Morel-Bokossa-Neerchal (MBN) bias-corrected Cluster-Robust Standard Error (CRSE), and covariates for study site and time (i.e. trial number). The estimated working correlation was $\widehat{\text{$\alpha$}} = .54$.\\
$^{**}p < 0.01$

\end{table}

\begin{figure}[htbp]

\caption{Boxplots of compliances at each possible score by trial.}
\label{fig:trust-compliance-by-score}
{
    \setkeys{Gin}{alt={Boxplots show an increasing trend of compliances for increasing performance in Trial 1 on the left and an increasing, jump down and increasing again trend for the same in Trial 2 on the left.}}
    \includegraphics[width=.75\linewidth]{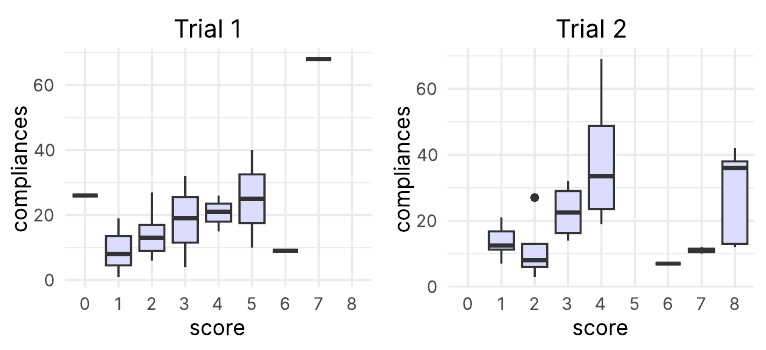}
}

\end{figure}

\subsection{RQ3: Does programming performance predict subsequent specific trust in an AIDE?}\label{sec:trust-rq3-does-programming-performance-predict-subsequent-specific-trust-in-an-aide-figure-1-c-to-a}

To investigate the connection in \Cref{fig:trust-cycle} from performance back to specific trust, we estimated a Gaussian GEE regression that included all four timepoints of specific trust scores. This allowed for the estimation of within-cluster correlation according to the autoregressive (AR1) structure supported by both trust research \cite{rodriguez_rodriguez_review_2023} and our \nameref{sec:trust-results-scale-validation-and-factor-analysis}. To differentiate the effects of a new experience from performance during that experience, we used a time factor covariate for whether the trust ratings were given pre- or post-trial. We then crossed this with score, keeping the product, but excluding the score-only predictor. Otherwise, the regression would nonsensically allow pre-trial trust to be predicted by a performance that had not yet happened (lookahead bias). Appropriately, the remaining product term goes to zero for pre-trial observations, with any systematic differences between pre- and post-trial ratings being captured by the time term. As shown in \Cref{fig:trust-perf-dist}, performance trended higher on trial two, so to avoid multicollinearity, no trial term (1 v. 2) was included. \Cref{table:trust-rq3-result} shows the results of the coefficient Wald tests. There was a small but statistically significant positive relationship between performance and subsequent trust, with each additional performance point (0-8) increasing trust by a fraction of a trust point (0-10). The significant covariates indicate generally higher trust ratings at VSU and decreases in trust following new AIDE experiences while controlling for performance. The interpretation is that it takes a score of \(\frac{.46}{.17} \approx\) 3 to maintain trust, with a lower or higher score changing trust accordingly. The \nameref{sec:trust-conclusion-sensitivity-analysis} revealed minor sensitivity. Neither the removal of two influential clusters nor the inclusion of a trial number predictor had major impacts. Mean trust aggregation and performance dichotomization affected the significance of covariates but not the score term. The estimated working correlation of \(\widehat{\text{$\alpha$}}\text{ = .77}\) supports the suspected autoregressive dependence structure.

\begin{table}[htbp]
\caption{Autoregressive gaussian GEE estimates relating performance score to post-trial trust.}
\label{table:trust-rq3-result}
\begin{flushleft}
{
\begin{tabular}{l llll}
\toprule
\multicolumn{1}{c}{\textbf{Variable}} & 
\multicolumn{1}{c}{\textbf{\textit{B}}} & 
\multicolumn{1}{c}{\textbf{\textit{CRSE}}} & 
\multicolumn{2}{c}{\textbf{95\% CI}} \\

& & & \multicolumn{1}{c}{\textit{LL}} & \multicolumn{1}{c}{\textit{UL}} \\ 
\midrule

Constant & $\hat{\mu}=\text{6.32}^{***}$ & 0.29 & 5.74 & 6.89 \\
Site = VSU & 1.17$^{*}$ & 0.51 & 0.16 & 2.17 \\
Post-Trial = True & -0.46$^{*}$ & 0.22 & -0.88 & -0.03 \\
(Post-Trial = T) x Score  & 0.17$^{***}$ & 0.05 & 0.07 & 0.26 \\

\bottomrule
\end{tabular}
}
\end{flushleft}
\noindent
\emph{Note}. $k=27$ clusters (participants) of balanced size $n_i=4$ (before and after each of two trials per participant). We estimated the population-averaged relationship between the performance score and the computed factor score for post-trial trust using Generalized Estimating Equations (GEE) with a Gaussian link, autoregressive (AR-1) correlation structure, Morel-Bokossa-Neerchal (MBN) bias-corrected Cluster-Robust Standard Error (CRSE), and covariates for study site and whether a given trust score followed a new programming trial. Pre-trial trust ratings were included, so the interaction term estimated the effect of performance while counteracting lookahead bias. The estimated working correlation was $\widehat{\text{$\alpha$}} = .77$.\\
$^{*}p < 0.05$; $^{***}p < 0.001$

\end{table}

\begin{figure}[htbp]

\caption{Histogram of scores by trial.}
\label{fig:trust-perf-dist}
{
    \setkeys{Gin}{alt={A histogram with two categories of Trial 1 and Trial 2 shows most data around a score of 2, with Trial 1 data trailing off up to 8 and quickly down to 0 and with Trial 2 data having a second spike down from 8.}}
    \includegraphics[width=.75\linewidth]{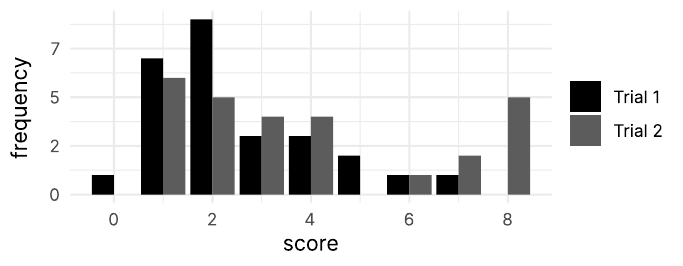}
}

\end{figure}

\section{Discussion}\label{sec:trust-discussion}

Overall, the results were inconclusive about a link from trust to AIDE compliance but did support relationships from compliance to performance and from performance to trust. Pre-interaction specific trust was not a significant predictor for AIDE compliance, but there was a positive relationship between compliance and performance. Performance was also positively correlated to post-interaction trust.

\subsection{RQ1: Does specific trust in an AIDE predict subsequent compliance?}\label{sec:trust-rq1-does-specific-trust-in-an-aide-predict-subsequent-compliance-figure-1-a-to-b-1}

We explored the relationship between pre-trial specific trust (1A and 2A in \Cref{fig:trust-timeline}) and the number of compliance events recorded during each trial. Here, we used trust as a predictor rather than a response in our GEE analysis. The negative sign of the estimated coefficient is opposite to the expectation from prior work \cite{patton_relationship_2024}, but the Wald test provided insufficient evidence to conclude whether the population-averaged rate of compliance events varied according to pre-trial specific trust. That is, the effect of trust on compliance was inconclusive. This outcome aligns to some recent findings that this effect, if present, is relatively small \cite{patton_relationship_2024,pharmer_transparent_2025}. The lack of alignment to earlier theory \cite{lee_trust_2004} and prior studies \cite{daly_task_2002,merritt_not_2008,wetzel_driver_2005} should be interpreted as a motivation for additional research rather than complete evidence that compliance is unrelated to trust.

Our findings are perhaps best explained by the lack of risk associated with compliance in our study. Participants likely incurred very little marginal cost for each additional compliance. Prather et al. discussed this "no downside" \cite{prather_its_2024} phenomenon in a qualitative study of novices using Copilot for the first time without time pressure. It is also consistent with the theory of automation bias \cite{mosier_automation_1999}. In our study, as easily as participants accepted code, they could evaluate it with pre-written tests and undo any changes they wished. Thus, a downside to compliance was the time cost of investing attention in Copilot's suggestions. Even a participant with low trust in Copilot could conceivably accept its suggestions as long as the potential upside was worth the time cost. In a scenario like that reported by Ross et al. \cite{ross_programmers_2023} where the performance of code could not be assessed objectively, users might comply less due to the momentary uncertainty about the marginal value of accepting a suggestion. Fundamentally, trust in automation is characterized by uncertainty about its true reliability \cite{lee_trust_2004}, such that feedback mechanisms aligned with the user's goals (e.g. the very same unit tests used to measure the performance they are striving toward) can downplay the role of trust in determining compliance behaviors. In \cite{lee_trust_1992}, participants' strong tendency for self-reliance was only overcome when the operator's trust in automation was very high compared to their self-confidence. They were not incentivized to risk the downsides of the automated machinery controllers performing poorly unless they did not think they could manage them manually. By comparison, our participants were incentivized to comply with the AIDE because, aside from time cost, they only stood to gain from its help.

The inconclusive relationship between trust and compliance might also be explained by the limitations of our compliance measure. While the kind of compliance we measured might not relate to trust, some other kind might. As described in the \nameref{sec:trust-method} section, we counted a compliance event whenever a user used a built-in compliance feature to accept code from Copilot. This included tab completions and clickable code insertion and editing tools. However, if participants complied with Copilot's suggestions by manually typing them out or copy-and-pasting suggested code, our data collection system did not record this. Whereas "slow accepting" \cite{prather_its_2024} suggestions by typing them out character-by-character, which was not tracked, was an investment of time perhaps risky enough to be done on the basis of trust, the counted compliances---those triggered by pressing the tab key or clicking a button---were presumably faster, less risky, and thus maybe less related to trust. Finally, compliance is easily reversible by discarding accepted code. Sabouri et al. \cite{sabouri_trust_2025} observed 10 developers using AI assistance and found they kept only about half of the code they first accepted. Future work should investigate whether discounting compliances according to reversals affects the relationship to initial trust.

\subsection{RQ2: Does compliance predict performance when coding with an AIDE?}\label{sec:trust-rq2-does-compliance-predict-performance-when-coding-with-an-aide-figure-1-b-to-c-1}

We next explored the relationship between compliance and performance and found a significant positive relationship between compliance and performance scoring rate, with each additional compliance event (range 1 to 69) corresponding to about a 2\% performance increase. Prior work suggested that this could be the case if the AIDE was sufficiently reliable to align with how much participants chose to use it \cite{klingbeil_trust_2024,lee_trust_2004,mcguirl_supporting_2006,passi_overreliance_2022}. On such simple tasks, Copilot should have been highly reliable \cite{finnie-ansley_my_2023}. As discussed above, for reliability in this scenario to be sufficient, suggestions needed to contain enough useful insights to help participants produce correct code without costing too much time. Even as participants were expected to develop their own strategies during the experiment, those who used the AIDE more seemed to have benefited from doing so. This finding emphasizes the risk of automation disuse (under-reliance), which has often been eclipsed in existing literature by the risk of misuse (over-reliance) \cite{kazemitabaar_how_2024,parasuraman_humans_1997,prather_its_2024,ross_programmers_2023,vaithilingam_expectation_2022}. When someone underuses AI-generated content, the performance of the human-AI team can suffer as a result, regardless of the AI's true reliability for the task at hand.

\subsection{RQ3: Does programming performance predict subsequent specific trust in an AIDE?}\label{sec:trust-rq3-does-programming-performance-predict-subsequent-specific-trust-in-an-aide-figure-1-c-to-a-1}

Prior work suggested that if participants had a sense of how they were performing with the AIDE, that their trust would follow suit \cite{dzindolet_role_2003,lee_trust_1992,lee_trust_2004,merritt_not_2008,wetzel_driver_2005,yang_toward_2023}, and our results confirmed this expectation. Here, we used trust as a response rather than a predictor in our GEE analysis. We found a significant positive relationship between performance score and post-trial trust ratings while accounting for the autoregressive nature of trust. The practical significance of the observed effect is small at less than 1\% of the trust scale's range per additional point scored of eight. We know from our previous quantitative analysis that participants from UVA performed significantly better on their second trial with the AIDE compared to the first \cite{gardella_performance_2024}. While the GEE approach accounted for the correlation of participants' trust ratings over time, a possible explanation for the relatively small observed effect of performance is the idea that trust has "inertia" \cite{hu_computational_2019,lee_trust_1992}. With trust "inertia," participants who had trust-building experiences in trial one might not have fully adjusted their trust until later. Especially if participants felt their trust was betrayed in some way in trial one, either the accumulation of positive experiences from both trials or an inertia of positive experiences from trial one over time may have been needed before their trust would increase.

\subsection{Implications}\label{sec:trust-implications}

Understanding trust is useful for an engineer if: (a) it affects a metric of importance to their organization and (b) it can be changed through a feasible intervention. We focused on the former by assessing how a novice's trust in an AIDE affects their compliance and performance on programming tasks. With the low-risk conditions in our study and the limitations of our compliance measure, we could not conclude whether participants' pre-trial specific trust had any meaningful impact on compliance. Therefore, when considering novices using AIDEs in performance-oriented task environments, our results did not suggest whether measures to manipulate trust will be useful in changing compliance behavior. However, compliance was positively associated with performance, so if maximal performance is the goal, training should focus on discouraging disuse in addition to discouraging misuse. Trust is a multifaceted attitude that may relate to other metrics of interest, such as workload and emotion, that were not studied here \cite{daly_task_2002,merritt_affective_2011}. Our findings indicate that novices trusted the AIDE more when it helped them perform well. To improve novices' trust outside of training materials, engineers can also focus on the ongoing effort toward making AIDEs themselves more reliable and trustworthy \cite{bolton_trust_2022,horne_pwnpilot_2023}.

\subsection{Limitations}\label{sec:trust-limitations}

The primary limitations of our study were how we measured compliance and the design choice to give participants a performance feedback tool (unit tests). As described in the \nameref{sec:trust-method}, we were only able to measure compliance events when participants used built-in compliance features, so there is a possibility that trust was related to copy-paste or typing initiated compliances in a way we were not able to observe. Furthermore, participants' easy access to performance feedback limited the all-important influence of uncertainty on specific trust \cite{hoff_trust_2015,lee_trust_2004}. While sample size and generalizability are always challenges, we were able to support some theories about trust derived from prior human factors research. The higher trust ratings and lower compliance rates at VSU compared to UVA emphasize the need for caution when generalizing over time and into varied contexts. The limited sample size split between UVA and VSU and the inconsistencies in software, training, etc. should motivate extreme caution when interpreting differences across sites. Because we did not measure learning, we cannot make claims regarding the role of trust in effective learning, but we encourage future work in this area. Finally, given practitioners may intend to use this work to inform trust-targeted interventions, we should emphasize that our findings rely only on natural variations in trust that happen during AIDE interactions. By contrast, when trust is intentionally manipulated, it may vary to a greater extent, uncovering an otherwise hidden relationship with compliance \cite{biros_effect_2003,spain_effects_2009}.

\section{Conclusion}\label{conclusion-1}

Conventional wisdom advises placing appropriate levels of trust in automation to avoid misuse or disuse. There is a contextually varied gap between what automation can do and whether humans think they can rely on it. In theory, this determines whether they do rely on it, creating a feedback loop \cite{lee_trust_2004}. We assessed this theory in the context of novice programmers using GitHub Copilot under time pressure. We measured specific trust in Copilot, compliance to its suggestions, and programming performance to test the validity of theory in our context. We failed to find a conclusive link from compliance to performance. However, compliance and performance were positively associated, and higher performance led to increased trust. These findings indicate that trust is related to performance and compliance but that manipulating trust to improve the performance of hybrid intelligence teams is a questionable approach that warrants further study. Practitioners should seek out more direct interventions for their metrics of interest and continue to make AIDEs more trustworthy programming companions.

\section{Key Points}\label{sec:trust-key-points}

\begin{itemize}
\item
  27 novice programmers at two research sites programmed with AI (GitHub Copilot) under time pressure for two 20-minute trials.
\item
  Trust in GitHub Copilot was measured four times with an adaptation of the Muir scale, performance was measured based on number of tasks completed per trial, and compliance was measured when Copilot's built-in compliance features were used.
\item
  Pre-trial trust failed to predict measures of compliance.
\item
  Performance was positively associated with AI compliance and post-trial trust.
\end{itemize}

\printbibliography
\section{Biographies}

\subsection{\texorpdfstring{Nicholas Gardella }{Nicholas Gardella }}
\noindent
Nicholas received the Master of Engineering degree in Systems Engineering from the University of Virginia in Charlottesville, VA, USA in 2023. He expects to receive the Ph.D. degree in Systems Engineering from the University of Virginia in 2026. He passed his dissertation defense and is currently a Ph.D. Candidate in the Department of Systems and Information Engineering, University of Virginia, Charlottesville, VA, USA.

\subsection{Matthew L. Bolton}
\noindent
Matthew received the Ph.D. degree from the University of Virginia in Charlottesville, VA, USA in 2010. He is currently an Associate Professor in the Department of Systems and Information Engineering, University of Virginia, Charlottesville, VA, USA.

\subsection{Sara L. Riggs}
\noindent
Sara received the Ph.D. degree in Industrial and Operations Engineering from the University of Michigan, Ann Arbor, MI, USA, in 2014. She is currently an Associate Professor in the Department of Systems and Information Engineering, University of Virginia, Charlottesville, VA, USA.

\end{document}

